\documentclass[sigconf]{acmart}
\AtBeginDocument{%
  \providecommand\BibTeX{{%
    \normalfont B\kern-0.5em{\scshape i\kern-0.25em b}\kern-0.8em\TeX}}}

\acmConference[CHI '23]{CHI '23: CHI Conference on Human Factors in Computing Systems}{April 23 -- 28, 2023}{Hamburg, Germany}
\acmBooktitle{CHI '23: CHI Conference on Human Factors in Computing Systems, April 23 -- 28, 2023, Hamburg, Germany}
\acmPrice{15.00}
\acmISBN{978-1-4503-XXXX-X/18/06}

\usepackage[linesnumbered,ruled]{algorithm2e}
\copyrightyear{2023} 
\acmYear{2023} 
\setcopyright{rightsretained} 
\acmConference[CHI '23]{Proceedings of the 2023 CHI Conference on Human Factors in Computing Systems}{April 23--28, 2023}{Hamburg, Germany}
\acmBooktitle{Proceedings of the 2023 CHI Conference on Human Factors in Computing Systems (CHI '23), April 23--28, 2023, Hamburg, Germany}\acmDOI{10.1145/3544548.3581158}
\acmISBN{978-1-4503-9421-5/23/04}
\begin{document}

\title{WebUI: A Dataset for Enhancing Visual UI Understanding with Web Semantics}

\author{Jason Wu}
\affiliation{%
  \institution{HCI Institute, Carnegie Mellon University}
  \city{Pittsburgh}
  \state{PA}
  \country{USA}}
\email{jsonwu@cmu.edu}
\author{Siyan Wang}
\affiliation{%
  \institution{Wellesley College}
  \city{Wellesley}
  \state{MA}
  \country{USA}}
\email{sw1@wellesley.edu}
\author{Siman Shen}
\affiliation{%
  \institution{Grinnell College}
  \city{Grinnell}
  \state{IA}
  \country{USA}}
\email{shenlisa@grinnell.edu}
\author{Yi-Hao Peng}
\affiliation{%
  \institution{HCI Institute, Carnegie Mellon University}
  \city{Pittsburgh}
  \state{PA}
  \country{USA}}
\email{yihaop@cs.cmu.edu}
\author{Jeffrey Nichols}
\affiliation{%
  \institution{Snooty Bird LLC}
  \country{USA}}
\email{jwnichls@gmail.com}
\author{Jeffrey P. Bigham}
\affiliation{%
  \institution{HCI Institute, Carnegie Mellon University}
  \city{Pittsburgh}
  \state{PA}
  \country{USA}}
\email{jbigham@cs.cmu.edu}
\renewcommand{\shortauthors}{Wu et al.}

\begin{abstract}
Modeling user interfaces (UIs) from visual information allows systems to make inferences about the functionality and semantics needed to support use cases in accessibility, app automation, and testing. Current datasets for training machine learning models are limited in size due to the costly and time-consuming process of manually collecting and annotating UIs. We crawled the web to construct WebUI, a large dataset of 400,000 rendered web pages associated with automatically extracted metadata.
We analyze the composition of WebUI and show that while automatically extracted data is noisy, most examples meet basic criteria for visual UI modeling.
We applied several strategies for incorporating semantics found in web pages to increase the performance of visual UI understanding models in the mobile domain, where less labeled data is available: \textit{(i)} element detection, \textit{(ii)} screen classification and \textit{(iii)} \textcolor{black}{screen similarity}.
\end{abstract}


\maketitle

\section{Introduction}
Computational modeling of user interfaces (UIs) allows us to understand design decisions \cite{deka2017rico,kumar2013webzeitgeist},  improve their accessibility \cite{zhang2021screen}, and automate their usage \cite{li2020mapping,li2017sugilite,burns2022interactive}. Often, \textcolor{black}{these systems must interact with UIs} in environments with incomplete or missing metadata (\textit{e.g.,} mobile apps authored with inaccessible UI toolkits). This presents many challenges since it necessitates that they reliably identify and reason about the functionality of the UI to support downstream applications. Visual modeling of UIs, which has shown to be a promising solution, predicts information directly from a screenshot using machine learning models and introduces no additional dependencies. 

Building the datasets needed to train accurate visual models involves collecting a large number of screenshots paired with their underlying semantic or structural representations. Recent efforts to collect datasets \cite{deka2017rico, zhang2021screen} for data-driven modeling have focused on mobile apps, which are typically manually crawled and annotated by crowdworkers since they are often difficult to automate. This process is both time-consuming and expensive — prior work has estimated that collecting a dataset of 72,000 app screens from 10,000 apps took 5 months and cost \$20,000 \cite{deka2017rico}. Because of this, datasets for visual UI modeling are limited in size and can be prohibitively expensive to keep updated.

The web \textcolor{black}{presents a possible solution to UI data scarcity since web pages are} a promising source of data to bootstrap and enhance visual UI understanding. In contrast to mobile UIs, web UIs (\textit{i.e.,} web pages) are much easier to crawl since they are authored in a unified parsable language (\textit{i.e.,} HTML) that typically exposes semantics (\textit{e.g.,} links and listeners) necessary for automated navigation. The same web page can also be viewed in many different viewports and display settings, which makes it possible to collect a large dataset of UIs rendered on a variety of devices (\textit{e.g.,} a smartphone or tablet). In addition, web browsers offer several facilities to extract visual, semantic, and stylistic information programmatically. In particular, web conventions, such as the semantic HTML and the ARIA initiatives, while not always adopted, constitute a large, if potentially noisy, source of annotations for UI elements. Finally, the web offers a virtually unlimited supply of data and has already been employed as a data source for large-scale machine learning \cite{gao2020pile,yalniz2019billion,xie2020self}. We explore the possibility of automatically collecting and labeling a large dataset of web UIs to support visual UI modeling in other domains (e.g., mobile). Compared to previous web datasets \cite{kumar2013webzeitgeist}, our dataset is much larger, more recent, and contains semantic information needed to support common visual UI understanding tasks.

\textcolor{black}{In this paper, we show that a large dataset of automatically collected web pages can improve the performance of visual UI Understanding models through transfer learning techniques, and we verify this phenomenon for three tasks.} We first describe the platform that we built to crawl websites automatically and scrape relevant visual, semantic, and style data. Our crawler visited a total of approximately 400,000 web pages using different simulated devices. WebUI, the resulting dataset is an order of magnitude larger than other publicly available datasets \cite{kumar2013webzeitgeist}. Next, we analyzed our dataset’s composition and estimated data quality using several automated metrics: \textit{(i)} element size, \textit{(ii)} element occlusion, and \textit{(iii)} layout responsiveness. We found that most websites met basic criteria for visual UI modeling. Finally, we propose a framework for incorporating web semantics to enhance the performance of existing visual UI understanding approaches. We apply it to three tasks in the literature: \textit{(i)} element detection, \textit{(ii)} screen classification and \textit{(iii)} video \textcolor{black}{screen similarity} and show that incorporating web data improves performance in other target domains, even when labels are unavailable. 

To summarize, our paper makes the following contributions:

\begin{enumerate}
    \item The WebUI dataset, which consists of 400,000 web pages each accessed with multiple simulated devices. We collected WebUI using automated web crawling and automatically associated web pages with visual, semantic, and stylistic information that can generalize to UIs of other platforms.
    \item An analyis of the composition and quality of examples in WebUI for visual UI modeling in terms of \textit{(i)} element size, \textit{(ii)} element occlusion, and \textit{(iii)} website layout responsiveness.
    \item A demonstration of the usefulness of the WebUI dataset through three applications from the literature: \textit{(i)} element detection, \textit{(ii)} screen classification and \textit{(iii)} \textcolor{black}{screen similarity}. We show that incorporating web data can lead to performance improvements when used in a \textit{transfer learning} setting, and we verified its improvement for our three tasks.
    We envision that similar approaches can be used for other tasks common in visual UI understanding. Furthermore, we show that models trained on only web data can often be directly applied to other domains (\textit{e.g.,} Android app screens). 
\end{enumerate}
All code, models, and data will be released to the public to encourage further research in this area.

\section{Related Work}
\subsection{Datasets for UI Modeling}
    There have been several datasets collected to support UI modeling, mostly in the mobile domain. Several datasets have been collected to support training specialized models \textcolor{black}{ \cite{he2021actionbert,moran2018machine,sermuga2021synz}}
    . The AMP dataset consists of 77k screens from 4,068 iOS apps and was originally used to train Screen Recognition, an enhanced screen reader \cite{zhang2021screen}, but has also been extended with additional pairwise annotations to support automated crawling applications \cite{feiz2022understanding}.

The largest publicly available dataset Rico, which consists of 72K app screens from 9.7K Android apps, was collected using a combination of automated and human crawling \cite{deka2017rico}. It captures aspects of user interfaces that are static (e.g., app screenshots) and dynamic (e.g., animations and user interaction traces). Rico has served as the primary source of data for much UI understanding research and it has been extended and re-labeled to support many downstream applications, such as natural language interaction \cite{wang2021screen2words,li2020mapping,burns2022interactive} and UI retrieval for design \cite{deka2017rico,bunian2021vins}.

Nevertheless, Rico has several weaknesses \cite{deka2021early}. Several works have identified labeling errors and noise (e.g., nodes in the view hierarchy do not match up with the screenshot). To this end, efforts have been made to repair and filter examples. Enrico first randomly sampled 10,000 examples from Rico then cleaned and provided additional annotations for 1460 of them \cite{leiva2020enrico}. The VINS dataset \cite{bunian2021vins} is a dataset for UI element detection that was created by collecting and manually taking screenshots from several sources, including Rico. The Clay dataset (60K app screens) was generated by denoising Rico through a pipeline of automated machine learning models and human annotators to provide element labels \cite{li2022learning}. Rico and other manually annotated datasets are expensive to create and update, and thus, models trained on them may exhibit degraded performance on newer design guidelines (e.g., Material Design is an updated design look for Android). For example, Rico was collected in early 2017 and has yet to see any update. Finally, many of these datasets focus on one particular platform (e.g., mobile phone) and therefore may learn visual patterns specific to the screen dimensions. For example, “hamburger menus” are usually used in mobile apps while desktop apps may use navigation bars.

In our work, we scrape the web for examples of UIs, which addresses some drawbacks (high cost, difficult to update, device-dependent) of current datasets but not others (dataset noise). The closest to our work is Webzeitgeist~\cite{kumar2013webzeitgeist}, which also used automated crawling to mine the design of web pages. To support design mining and machine learning applications, Webzeitgeist crawled 103,744 webpages and associated web elements with extracted properties such as HTML tag, size, font, and color. This work is primarily used for data-driven design applications and does not attempt to transfer semantics to other domains. We also collect multiple views of each website and query the browser for accessibility metadata, which can further facilitate UI modeling applications.

\subsection{Applications of UI Datasets}
Applications that operate and improve existing UIs must reliably identify their composition and functionality.
Originally, many relied on pixel-based or heuristic matching \cite{dixon2010prefab,yeh2009sikuli,outspoken,autoit}.
The introduction of large UI datasets, such as those previously discussed, have provided the opportunity to learn more robust computational models, especially those from visual data.
The goal of this paper is to improve the performance of these computational models by leveraging a large body of web data and its associated semantics.
There have been many efforts to learn the semantics of UIs \cite{liu2018learning,wu2021screen,wang2021screen2words}. In this paper, we focus on three modeling tasks at the \textit{(i)} element (element detection), \textit{(ii)} screen (screen classification), and \textit{(iii)} app-level (\textcolor{black}{screen similarity}).

Element detection identifies the location and type of UI widgets from a screenshot and has applications in accessibility metadata repair \cite{zhang2021screen}, design search \cite{bunian2021vins}, and software testing \cite{xie2020uied, chen2020object}.
Labeled datasets for element detection exist \cite{bunian2021vins,li2022learning,zhang2021screen,deka2017rico}; however they are quite small compared to other datasets for object detection \cite{lin2014microsoft} which contain an order of magnitude more examples (330K).
We found that incorporating our web UI dataset (400K examples) in a pre-training phase led to performance benefits.
Other work involves modeling UIs at a higher level (\textit{e.g.,} screen-level) to reason about the design categorization \cite{leiva2020enrico} and purpose \cite{wang2021screen2words} of a screen. 
Similarly, datasets with screen-level annotations of UIs are much smaller than others used in the CV literature \cite{deng2009imagenet} so we used additional web data to improve accuracy.
Finally, we investigated \textcolor{black}{screen similarity}, a task that reasons about multiple UI inputs (\textit{e.g.,} frames of a video recording), where no publicly available labeled data exists.
We found that models trained on related web semantics (\textit{e.g.,} URL similarity) were able to successfully generalize to mobile screens.
In summary, our paper shows that applying examples from the web and relevant machine learning techniques can improve the performance of computational models that depend on UI data.

\subsection{Related Machine Learning Approaches}
We briefly introduce and summarize three machine learning approaches that we apply in our paper. Broadly, they fall under a body of research known as “transfer learning” which uses knowledge from learning one task (e.g., web pages) to improve performance on another (e.g., mobile app screens).

Inductive transfer learning is a technique that improves model performance by first “pre-training” a model on a related task, typically where a lot of data is available \cite{pan2009survey}. Once the model converges on the first task, its weights are used as a starting point when training on the target task. Labeled data is required for both the source and target domains, although it is possible that there are fewer target examples.

In some cases, labeled data are missing for either the source or target domains. If source labels are unavailable, semi-supervised learning (SSL) can be applied to take advantage of unlabeled data to improve performance \cite{chapelle2009semi}. For example, WebUI doesn’t contain any labels for screen type (e.g., login screen, register screen), but we’d like to use it to improve prediction accuracy on a small number of annotated Android app screens. In our work, we apply a form of SSL known as “self-learning” \cite{chapelle2009semi}, where a UI classification model iteratively improves its performance by generating pseudo-labels for an unlabeled dataset, then re-training itself using high-confidence samples.

Finally, to support use-cases where target labels are unavailable, we apply unsupervised domain adaptation (UDA) \cite{ganin2016domain}. In many cases, visual UI models trained on web data can be directly used on any screenshot (including Android and iOS apps), and UDA improves the performance and robustness of models to domain changes. This type of knowledge transfer is particularly interesting because it enables us to explore the feasibility of new UI understanding tasks (without manually annotating a large number of examples) and bring some benefits of web semantics (e.g., semantic HTML) to other platforms.

\section{WebUI Dataset}
We introduce the WebUI dataset, which we construct and release to support UI modeling. The WebUI dataset is composed of 400,000 web pages automatically crawled from the web. We stored screenshots and corresponding metadata from the browser engine, which serve as annotations of UI element semantics. Because the collection process is highly automated, our final dataset is an order of magnitude larger than other publicly available ones (Figure \ref{fig:diagram_comparison}) and can be more easily updated over time.

In this section, we give an overview of our web crawling architecture, analyze the composition of our dataset, and provide evidence that it can support visual UI modeling for other platforms.
\subsection{Web UI Crawler}
\subsubsection{Crawling Architecture}
\begin{figure}[!]
    \centering
\includegraphics[trim={10pc 10pc 3pc 0},clip,width=15pc]{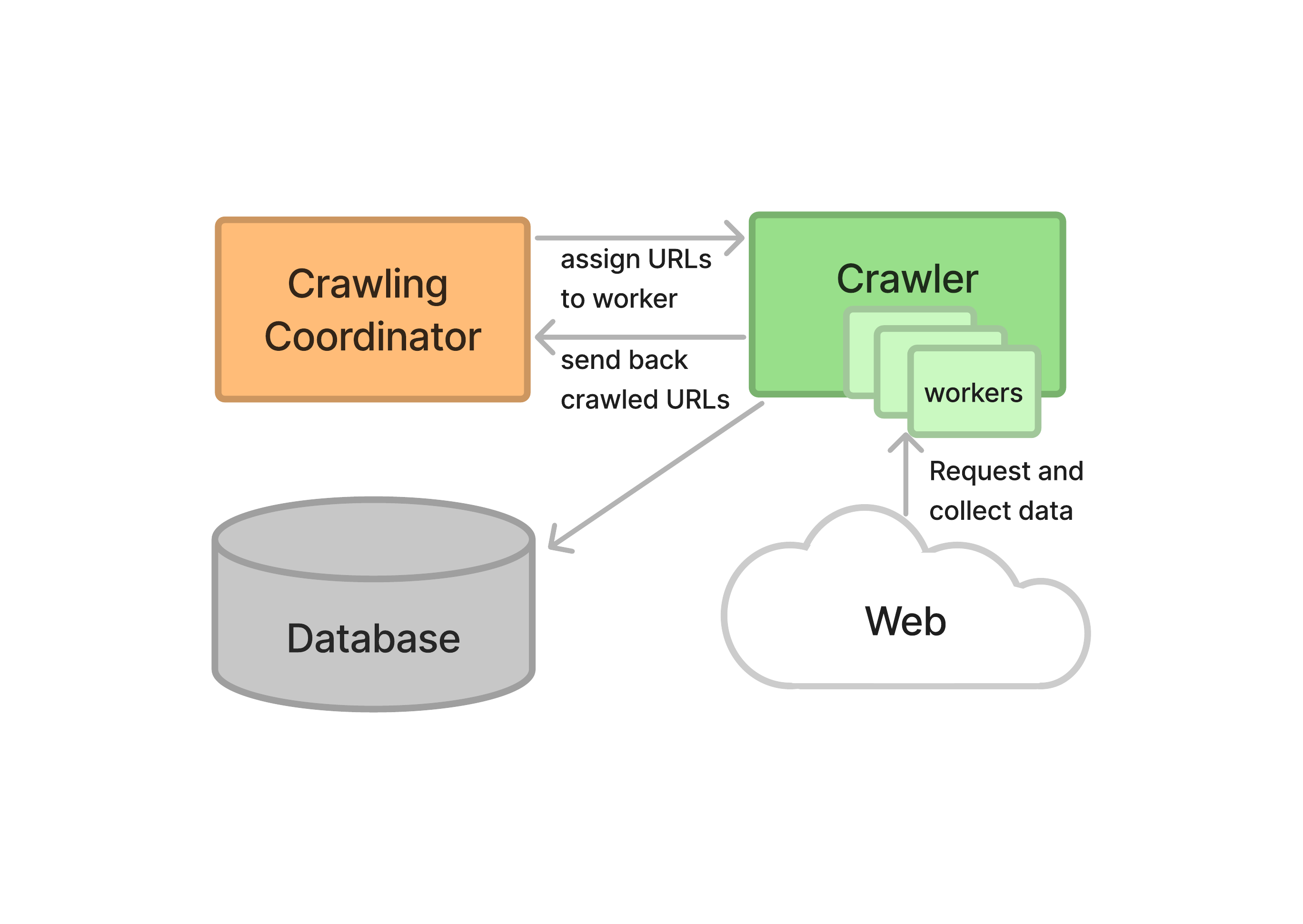}
\Description{Flow chart that shows our crawling architecture. The crawling coordinator assigns URLs to crawler workers. The crawler worker requests and collects data from the web and stores it in a database. The worker also sends back crawled URLs to the coordinator.}
    \caption{Overview of our crawling architecture. A \textit{crawling coordinator} contains a queue of URLs to crawl and assigns them to workers in a \textit{crawler pool}. Workers asynchronously process URLs by visiting them in a automated browser, scraping relevant metadata, then uploading them to a cloud database.}
    \label{fig:diagram_crawler}
\end{figure}
To collect our dataset, we implemented a parallelizable cloud-based web crawler. Our crawler consists of \textit{(i)} a crawling coordinator server that keeps track of visited and queued URLs, \textit{(ii)} a pool of crawler workers that scrapes URLs using a headless browser, and \textit{(iii)} a database service that stores uploaded artifacts from the workers. The crawler worker is implemented using a headless framework \cite{puppeteer} for interfacing with the Chrome browser. Each crawler worker repeatedly requests a URL from the coordinator server, which keeps global data structures for visited and upcoming URLs.  The crawler worker includes some simple heuristics to automatically dismiss certain types of popups (\textit{e.g.,} GDPR cookie warnings) to help it access page content. 

We seeded our coordinator using a list of websites that we hypothesized would lead to diverse examples of web pages (e.g., link aggregation websites and design blogs) and ones that we expected to have high-quality accessibility metadata (\textit{e.g.,} government websites). A full list of our seed websites can be found in the supplementary materials.

We explored several crawling policies and eventually settled on one that encourages diverse exploration by inversely weighting the probability of visiting a URL by its similarity to the visited set. For example, if the crawler previously visited \url{http://example.com/user/alpha}, it would be less likely to subsequently visit \url{http://example.com/user/beta}. We set a minimum probability so that it is possible to re-visit links to support additional types of analysis (e.g., temporal changes). The coordinator organizes upcoming (i.e., queued) URLs by their hostname, \textit{(i)} selects a hostname randomly with uniform probability, and then \textit{(ii)} selects a URL using its assigned probability. Empirically, we found this technique to be effective at avoiding crawler traps, which are websites that cause automated crawlers to get stuck in endless loops navigating within the same site. 
\subsubsection{Data Collected from a Web Page}
We used a pool of crawler workers to crawl web pages in parallel, and we visited each URL with multiple simulated devices. We collected several types of semantic information by querying the rendering and accessibility engine. We set a timeout limit of 6 minutes for each URL, so some web pages were not visited by all simulated devices.
\begin{figure}[!]
    \centering
\includegraphics[width=\linewidth]{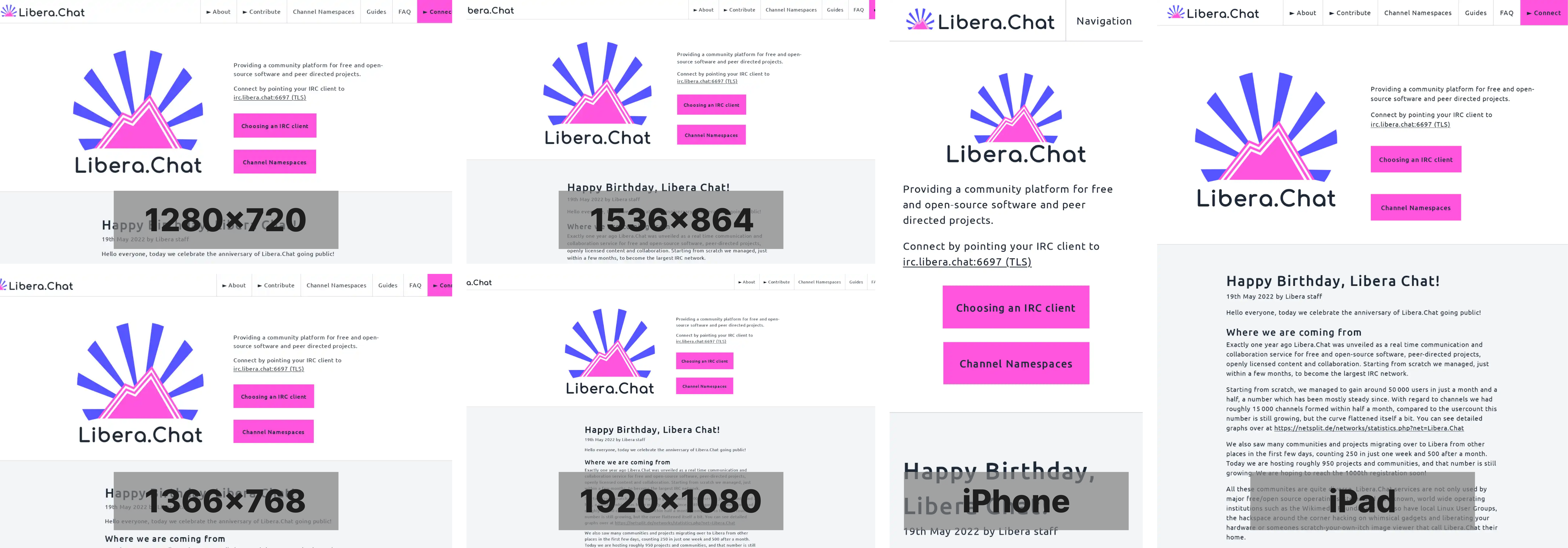}
\Description{A collage of screenshots illustrating how a web page appears differently when accessed using 6 different simulated devices.}
    \caption{Screenshots from a web page accessed using 6 different devices: 4 desktop resolutions, a smartphone, and a tablet. By requesting a responsive web page at different resolutions, we induce several layout variations (\textit{e.g.,} navigation and hero button).}
    \label{fig:diagram_devices}
\end{figure}

\textbf{Simulated Devices.} We sampled each web page with 6 simulated devices: 4 of the most common desktop resolutions \cite{browserstack}, a tablet, and a mobile phone. Devices are simulated by setting the browser window resolution and user agent to match the goal device, both of which may affect the page’s content and rendering.

\textbf{Screenshots.} Our crawler worker captured two types of screenshots (\textit{i.e.,} visual data) from websites. We captured a viewport screenshot, with fixed image dimensions, and a full-page screenshot, with variable height. Images were saved using lossy compression to save storage. While compression can introduce some artifacts, previous work \cite{dodge2016understanding} suggests that the effect on deep learning model performance is minimal.

\textbf{Accessibility Tree.} We used a browser automation library to query Chrome’s developer tools to retrieve an accessibility tree for each page \cite{chromeaxtree}. The accessibility tree is a tree-based representation of a web page that is shown to assistive technology, such as screen readers. The tree contains accessibility objects, which usually correspond to UI elements and can be queried for properties (\textit{e.g.,} clickability, headings).

Compared to the DOM tree, the accessibility tree is simplified by removing redundant nodes (e.g., <div> tags that are only used for styling) and automatically populated with semantic information via associated ARIA attributes or inferred from the node’s contents.
\textcolor{black}{The browser generates the accessibility tree using a combination of HTML tags, ARIA attributes, and event listeners (\textit{e.g.,} click handlers) to create a more consistent semantic representation of the UI. For instance, there are multiple ways to create a button (\textit{e.g.,} a styled \texttt{div}) and the accessibility tree is intended to unify all of these to a single \texttt{button} tag.}

\textbf{Layout and Computed Style.} For each element in the accessibility tree, we stored layout information from the rendering engine. Specifically, we retrieved 4 bounding boxes relevant to the “box model”: \textit{(i)} the content bounding box, \textit{(ii)} the padding bounding box, \textit{(iii)} the border bounding box, and \textit{(iv)} the margin bounding box. Each element was also associated with its computed style information, which included font size, background color and other CSS properties.
\subsection{Dataset Composition}
The WebUI dataset contains 400K web UIs captured over a period of 3 months and cost about \$500 to crawl.
We grouped web pages together by their domain name, then generated training (70\%), validation (10\%), and testing (20\%) splits. This ensured that similar pages from the same website must appear in the same split. We created \textcolor{black}{four} versions of the training dataset.
\textcolor{black}{Three of these splits were generated by randomly sampling a subset of the training split:} Web-7k, Web-70k, Web-350k. We chose 70k as a baseline size, since it is approximately the size of existing UI datasets \cite{zhang2021screen,deka2017rico}. 
\textcolor{black}{We also generated an additional split (Web-7k-Resampled) to provide a small, higher quality split for experimentation. Web-7k-Resampled was generated using a class-balancing sampling technique, and we removed screens with possible visual defects (\textit{e.g.,} very small, occluded, or invisible elements). More information about how this set was generated can be found in the appendix.}
The validation and test split was always kept the same.

\begin{figure}[!]
    \centering
\includegraphics[width=\linewidth]{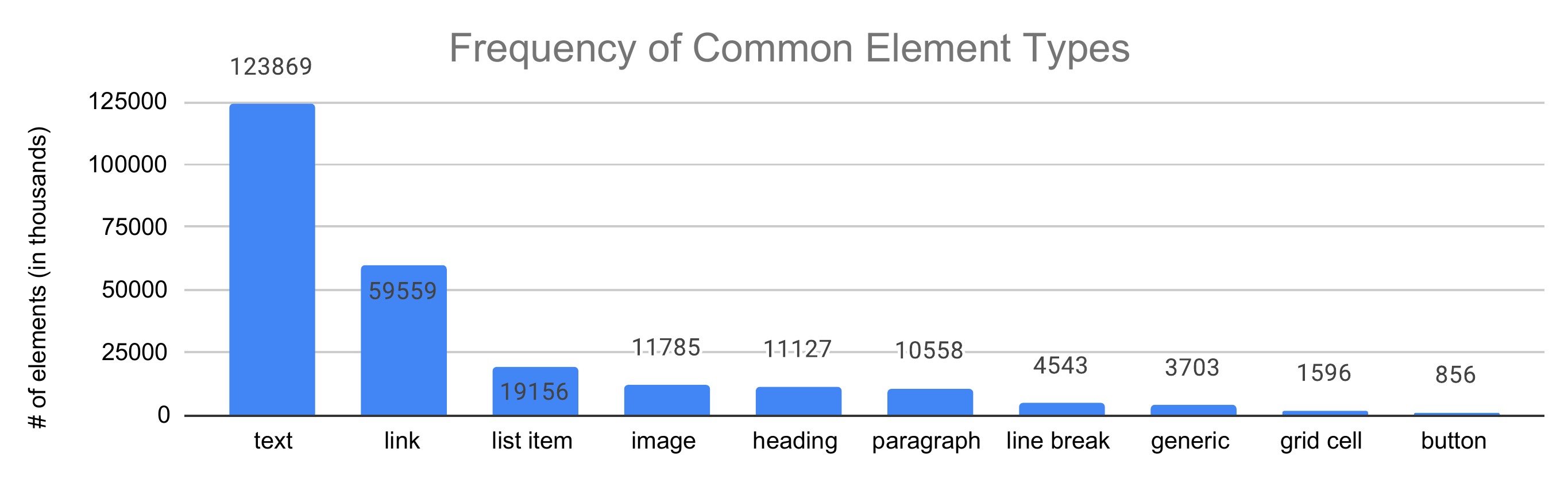}
\Description{Bar chart of the 11 most common element types in the WebUI dataset. The most common element type is text and the least common type is button. When sorted by frequency, there is a power-law distribution.}
\caption{\textcolor{black}{10} most common element types in the WebUI dataset. Element types are based on automatically computed roles, which are not mutually exclusive. Text is the most common type, but many types offer semantic information about what text is used for \textit{e.g,} a heading, paragraph or link.}
    \label{fig:diagram_composition}
\end{figure}
\subsubsection{Comparison to Existing Datasets}
\label{sec:dataset_comparison}
\begin{figure}[!]
    \centering
\includegraphics[width=\linewidth]{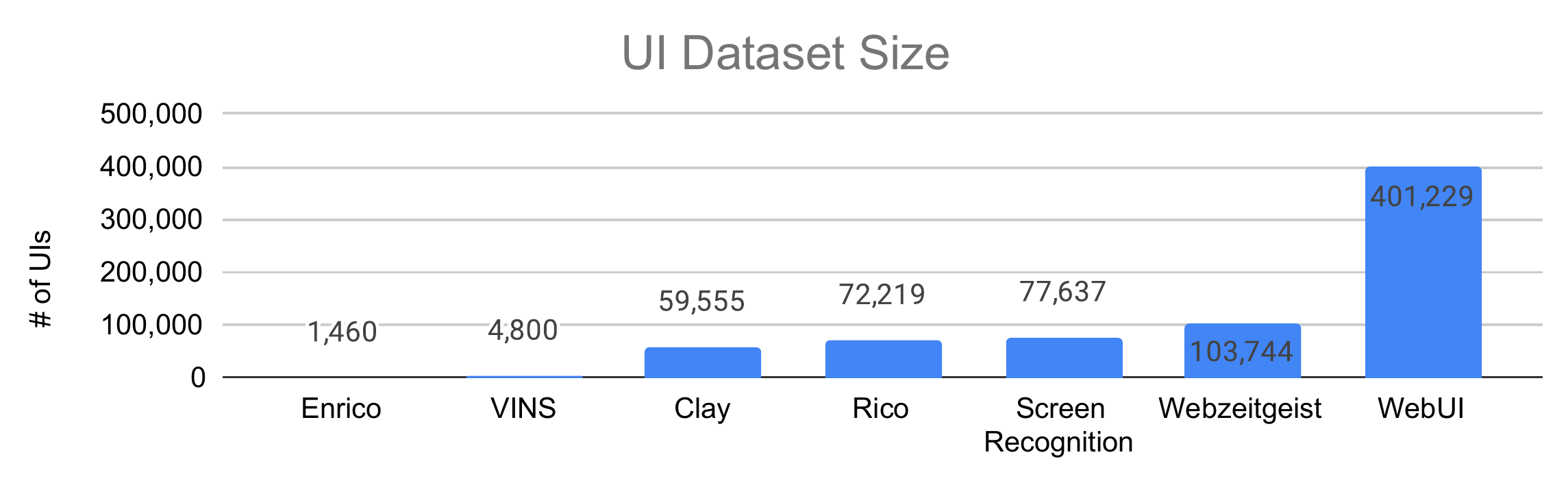}
\Description{Bar chart of 7 UI datasets and their size in number of UIs. WebUI is an order or magnitude larger than other publicly released datasets.}
    \caption{Comparison of WebUI to existing UI datasets. WebUI contains nearly 400,000 web pages and is nearly one order of magnitude larger than existing datasets available for download (Enrico, VINS, Clay, Rico). Each web page also contains multiple screenshots captured using 6 simulated devices.}
    \label{fig:diagram_comparison}
\end{figure}
WebUI is an order of magnitude larger than existing datasets used for UI understanding (Figure \ref{fig:diagram_comparison}) and provides rich semantic and style information not found in mobile datasets. WebUI focuses on the static properties of web pages and does not store page loading times or element animations.

We analyzed the makeup of web UIs and compared them to mobile UIs. The distribution of UI types (\textit{e.g.} Login, News, Search) in WebUI are also likely to be different than mobile data, since many web pages are primarily hypertext documents. We extracted elements from the accessibility tree and categorized them using their computed accessibility role and the role of any singleton parents. For example, a clickable image is created in HTML by surrounding an image (<img>) element with an anchor element  (<a>). Thus, it is possible for elements to be assigned to multiple classes. Figure \ref{fig:diagram_composition} shows the frequency of element types in our dataset. Similar to prior work \cite{zhang2021screen}, we find that text is the most common element in our dataset. However, we find limited overlap between the rest of the label set, possibly due to the nature of web data and the mutually exclusive nature of existing label sets. On average, there were 60 elements on a web UI, 30 of which were visible in the viewport. This is more than the number of elements on mobile app screens, which prior work estimated to be around 25 per screen, although this may in part be due to differences in segmentation (\textit{e.g.,} a single Rich Text Field on Android can contain differently formatted text while on HTML they would broken up into different tags). On average, there were also more clickable elements per web page (20 on web pages vs 15 ``interactable" elements on Android apps), likely due to the prevalence of hyperlinks on the web. 
\subsubsection{Dataset Quality}
Compared to manually labeled examples, automatically extracted annotations can contain errors that impact modeling performance. We conducted an analysis on a small, randomly sampled data from our dataset (1000 web pages). While there are numerous possible defects, we focus on three that we believe are most relevant to data quality: \textit{(i)} element size, \textit{(ii)} element occlusion, and \textit{(iii)} website responsiveness.
\textcolor{black}{Our analysis is primarily focused on quantifying possible defects but not reparing them. Previous work \cite{sermuga2021synz, li2022learning} has explored automated methods for correcting mismatched labels and occluded elements, and we expect the overall quality of WebUI could be improved if these were applied.}.

\textbf{Element Size.} Element size refers to the dimensions of an annotated object in an image. For example, if a bounding box annotation surrounds an object that is too small relative to the image resolution, it may be difficult for a model to identify the object. The average area of bounding boxes in our data is approximately $14000 px^2$, but this may have been influenced by short segments of text. The Web Content Accessibility Guidelines (WCAG) guideline for target size also recommends that interactable elements have a minimum size of 44 by 44 pixels, so that they can be easily selected by users. In our dataset, one third of interactable elements (\textit{e.g.,} elements tagged as links or button) were smaller than this threshold.

\textbf{Element Occlusion.} Element occlusion occurs when one object partially or completely covers another in a screenshot. Occluded elements are detrimental to visual modeling since they may represent targets that can be impossible to predict correctly. We quantified the occlusion rate by counting the number of screens with overlapping \textcolor{black}{leaf elements}. We found that 18\% of screens in our sampled split contained overlapping \textcolor{black}{leaf elements}. However, of the overlapping elements, only a third of them were occluded by more than 20\% of their total area.

\textbf{Responsive Websites.} Website responsiveness relates to how well a web page adapts to different screen viewports. Since we simulated multiple devices for each web page, responsive websites are likely to produce more variation in their layouts than unresponsive ones. To measure responsiveness, we automatically computed metrics included in the Chrome Lighthouse tool for estimating layout responsiveness: \textit{(i)} responsiveness of content width to window size and \textit{(ii)} the use of a viewport meta tag, which is needed for proper mobile rendering. From our analysis we found that 70\% and 80\% of processed web pages met the first, and second criteria, respectively. 

In summary, our analysis suggests that most web pages in our dataset meet some basic requirements for visual UI modeling.
Given the reliance of our data collection on extracted accessibility metadata, we expect high quality examples to adhere to good accessibility practices, such as those outlined by WCAG.
However, considering the inaccessibility of the web and that many criteria are difficult to verify automatically, we also expect many web pages to violate some of these criteria.
There are other desirable properties for dataset quality that we did not check, \textit{e.g.,} the accurate use of semantic HTML tags, ARIA tags, and tightness of element bounding boxes.
These properties were harder to verify automatically, since they require knowledge of developer intention and associated tasks.
In our analysis, we only attempt to identify possible defects, and we did not attempt to remove or repair samples. This could be a direction for future work to improve dataset quality \cite{chang2011associating,li2022learning}.
\section{Transferring Semantics from Web Data}
We hypothesized that web data is similar and relevant to modeling other types of UIs from their pixels. In this paper, we are specifically interested in the mobile domain, as mobile apps often lack metadata and can only be reliably understood from their visual appearance.
In many cases, manually-annotated mobile datasets are small, and in some cases, labels are completely unavailable.
We used transfer learning to apply our dataset to three existing tasks in the UI understanding literature: \textit{(i)} element detection, \textit{(ii)} screen classification, and \textit{(iii)} \textcolor{black}{screen similarity}. Table \ref{tab:strategies} shows downstream applications where UI understanding tasks can benefit from web data.
\textcolor{black}{Because each task contains different constraints (\textit{e.g.,} presence of labeled target data) it is difficult to apply a single strategy to serve all use-cases. For example, inductive transfer learning typically requires labels in both the pre-training and fine-tuning phase is impossible to apply to a setting where target labels are unavailable (\textit{e.g.,} screen similarity). We expect our three transfer learning strategies to be applicable to most future use-cases, since they span all combinations of labeled data availability (Table \ref{tab:strategies}).}
\begin{table*}[!]
\Description{Strategies for transferring semantics from web pages to other types of UIs organized by whether labeled data is available in the source web domain or in the target mobile domain. If both web and mobile data are available, we apply finetuning. If only mobile data is available, we apply semi-supervised learning, if only web data is available, we apply domain adaptation.}
\caption{Table of strategies for transferring semantics from web pages to other types of UIs. We explored scenarios where labeled data is missing in either domain by applying three strategies: \textit{(i)} finetuning, \textit{(ii)} semi-supervised learning, and \textit{(iii)} domain adaptation.}
\label{tab:strategies}
\begin{tabular}{@{}llll@{}}
\toprule
Approach          & Finetuning & Semi-supervised Learning & Domain Adaptation \\ \midrule
Application & Element Detection & Screen Classification & \textcolor{black}{Screen Similarity} \\ \midrule
Web (Source)    & Y          & N                        & Y                 \\
Mobile (Target) & Y          & Y                        & N                 \\ \bottomrule
\end{tabular}
\end{table*}

\subsection{Element Detection}
\begin{figure}[!htb]
    \centering
\includegraphics[trim={10pc 10pc 3pc 10pc},clip,width=20pc]{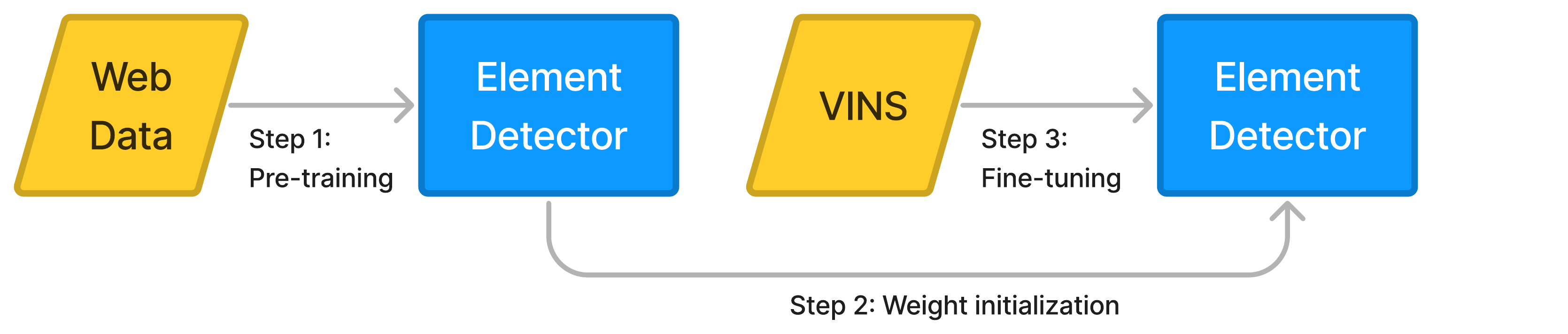}
\Description{Flow chart demonstrating how inductive transfer learning is applied to element detection. First web data is used to pre-train an element detector. The weights are used to initialize a new model, which is then fine-tuned on the down-stream VINS dataset.}
    \caption{We applied inductive transfer learning to improve the performance of a element detection model. First, we pre-trained the model on web pages to predict the location of nodes in the accessibility tree. Then, we used the weights of the web model to initialize the downstream model. Finally, we fine-tuned the downstream model on a smaller dataset consisting of mobile app screens.}
    \label{fig:diagram_finetune}
\end{figure}
Element detection requires a machine learning model to identify the locations and types of UI elements from a screenshot.
Often these models are based on object detection frameworks.

Element detection is an example of a task where labeled data is available in both the source and target domain (albeit fewer examples of mobile screens), so it is possible to employ inductive transfer learning. The WebUI dataset contains the locations of elements that we scraped from the website accessibility tree. Element types are inferred from the HTML tags and the ARIA labels \cite{chromeaxtree}.
We show that this training strategy results in improvements to element detection performance.

\subsubsection{Model Implementation}
We primarily followed the details provided by VINS \cite{bunian2021vins} to implement our element detection model.
The VINS dataset, which we used for training, is composed of 4800 annotated UI screenshots from various sources such as design wireframes, Android apps, and iOS apps.
Since the authors did not release official data splits, we randomly partitioned the data into training (70\%), validation (15\%), and testing (15\%) sets.
\textcolor{black}{This specific split ratio was chosen since it has been used in other UI modeling work \cite{wu2021screen}.}
The paper identifies 11 primary UI component classes; however the released raw dataset includes a total of 22 class labels.
For the extraneous labels, we either tried to merge them with the 11 primary labels (\textit{e.g.,} ``Remember Me" merged with ``Check Box") or assigned them to an ``Other" class (\textit{e.g.,} ``Map") if no good fit was found.
Instead of the SSD object detection model \cite{liu2016ssd} used by VINS, we opted to start from the more recent FCOS model architecture \cite{tian2019fcos}, since we found it was easier to modify to support multi-label training.
Previous element detection work \cite{bunian2021vins,zhang2021screen,chen2020object} trained models to assign one class label (\textit{e.g.,} Button, Text field) to each detected element in the screenshot. To take advantage of multiple, nested definitions of web elements in our dataset, we trained the object detection model to predict multiple labels for each bounding box.

Figure \ref{fig:diagram_finetune} illustrates the overall training process.
In the pre-training phase, the element detection model is trained on a split of the WebUI dataset.
\textcolor{black}{Due to cost and time constraints, we trained all element detection models for a maximum of 5 days. We also used early stopping on the validation metric to reduce the chance of overfitting.}
Afterwards, a specific part of the model was re-initialized (the object classification head) to match the number of classes in the VINS dataset before it was fine-tuned.
We found it difficult to modify the original SSD architecture to support the multi-label pre-training, so we only followed the original training from scratch procedure described in the paper as a baseline.
\subsubsection{Results}
Table \ref{tab:results_elementdetection} shows the performance of each model configuration on the VINS test set, and we show that our updated configurations lead to significant performance improvements.
\textcolor{black}{Our primary performance metric for this task was the mean average precision (mAP), which is a standard metric used for object detection models that takes into the accuracy of bounding box location (\textit{i.e.,} how closely the predicted box overlaps with ground truth) and classification (prediction of object type).
The mAP score is calculated by computing an individual average precision (AP) score for each possible element class (\textit{e.g.,} Text, Check Box), which represents the object detector's accuracy in detecting each object class. The AP scores are averaged to produce the mAP score.
We calculated the mAP score over classes that could be mapped to the original label set in the paper \cite{bunian2021vins} \textit{i.e.,} we excluded the ``Other" class where there was no clear mapping to the original set.
We calculated the un-weighted mean between class APs, which assigns equal importance to common and rare element types.}
Our best model configuration performed 0.14 better than the baseline in terms of mAP score.
While the largest source of improvement over the baseline configuration (SSD) came from the updated FCOS model architecture, our fine-tuning procedure contributed to gains as well. Specifically, we note that pre-training with more examples led to better performance (around 0.04 mAP).
\textcolor{black}{Depending on the downstream application of the element detection model, this improvement could lead to better user experience but would require further validation. For example, a screen reader \cite{zhang2021screen} does not require tight bounding boxes; however, it would benefit from detecting more (small) elements on the screen. Query-based design search \cite{bunian2021vins} could also retrieve more relevant examples.}

Although we followed the original training procedure as closely as possible, we were unable to reach the mAP score reported in the original VINS paper.
This can be attributed to \textit{(i)} our use of different randomized splits and \textit{(ii)} differences in mappings between class labels from the raw data to the 11 primary classes, which were not provided in the previously released code.
Nevertheless, since we used the same splits and class mappings across all of our model configurations, we expect the relative performance improvements to be consistent.

\begin{figure*}[!]
    \centering
\includegraphics[width=\textwidth]{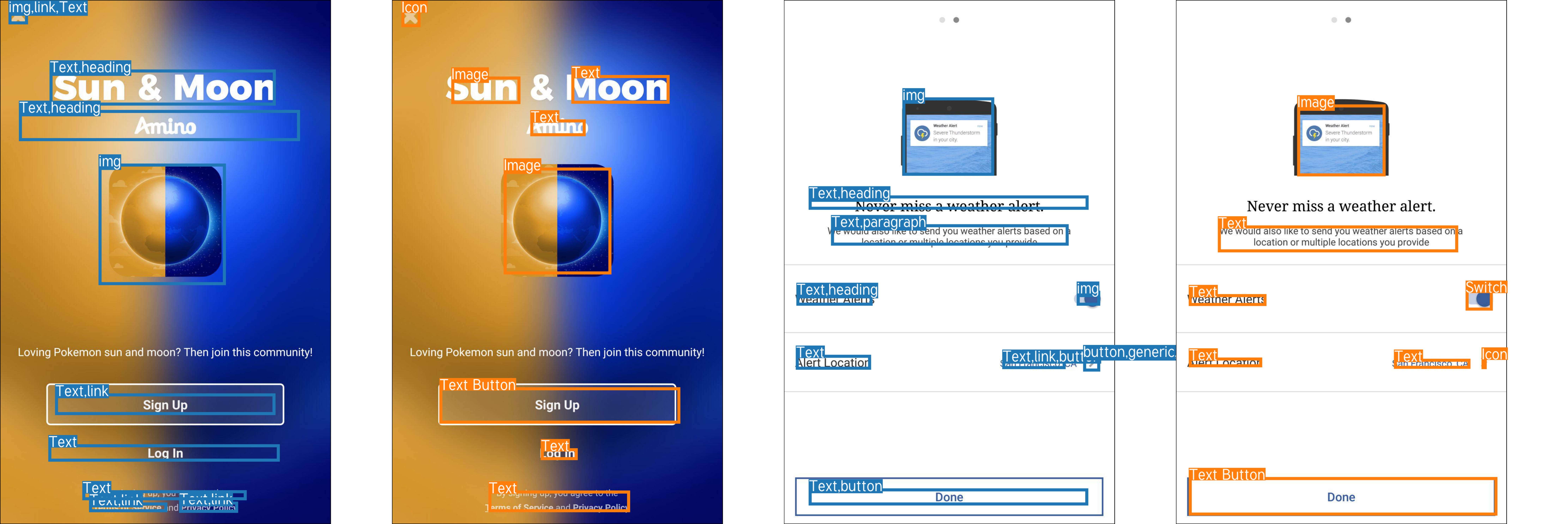}
\Description{2 app screens each processed by our web element detector and mobile element detector. Both screens contain correctly detected elements, although the web model’s output has less accurate bounding boxes. The labels generated by our web model contain additional tag predictions that offer additional semantics, like heading and link.}
    \caption{Output of our element detection models run on two app screens. In many cases, detections from our web-only model (Blue) coincide with ones from our fine-tuned model (Orange), which suggests some zero-shot transfer capabilities. Predicted tags from the web-only model also provide additional metadata corresponding to clickability (\texttt{link}) and heading prediction (\texttt{heading}); however, the predicted bounding boxes are often less tight than the fine-tuned model.}
    \label{fig:diagram_zsl}
\end{figure*}

We also investigated the zero-shot performance of element detectors trained only on web data (\textit{i.e.,} without fine-tuning).
It is difficult to compute performance quantitatively, since the label sets between the web and mobile datasets do not directly overlap. However, we provide qualitative evidence that zero-shot learning could be successful.
Figure \ref{fig:diagram_zsl} shows the output of a web model when run on mobile app screens from Rico. We conducted minimal preprocessing, such as cropping out the Android system notification bar and the navigation soft buttons.
In many cases, the web analogs of mobile text and image elements are detected accurately, which suggests that some element classes have consistent appearance across platforms.
Interestingly, some web classes such as links and headings are also detected in the image, which could be used to infer new semantics such as clickability \cite{swearngin2019modeling} and navigation landmarks.

\begin{table}[!]
\Description{Element detection performance of 5 model configurations. The baseline configuration achieves a mAP score of 0.6737 while our best configuration achieves a mAP score of 0.8115.}
\caption{Element detection performance (11 object classes) for different model configurations. Pre-training on more web screens led to better performance on mobile screens after fine-tuning.}
\label{tab:results_elementdetection}
\begin{tabular}{@{}ll@{}}
\toprule
Model Configuration         & mAP    \\ \midrule
SSD (Random Init.)            & 0.6737 \\
FCOS (Random Init.)           & 0.7739 \\
FCOS (Pre-trained on Web7k)   & 0.7877 \\
\textcolor{black}{FCOS (Pre-trained on Web7k-Resampled)}   & \textcolor{black}{0.7961} \\
FCOS (Pre-trained on Web70k)  & 0.7921 \\
FCOS (Pre-trained on Web350k) & 0.8115 \\ \bottomrule
\end{tabular}
\end{table}
\subsection{Screen Classification}
\begin{figure}[!htb]
    \centering
\includegraphics[trim={10pc 10pc 3pc 10pc},clip,width=20pc]{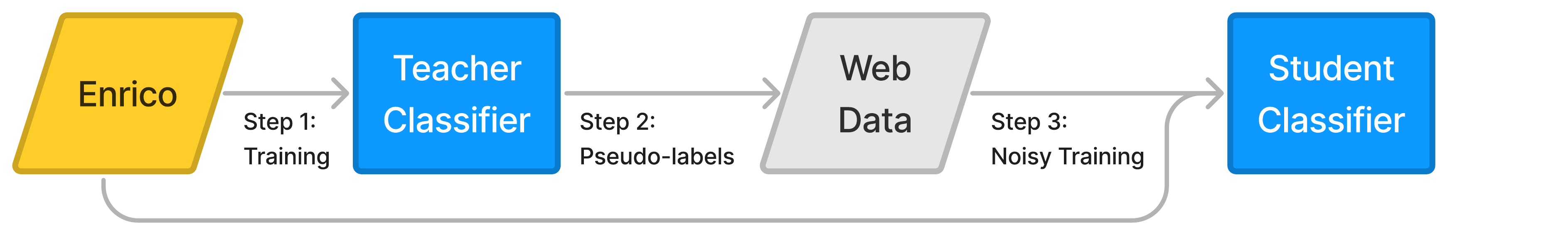}
\Description{Flow chart demonstrating how semi-supervised learning is applied to screen classification. First the Enrico mobile dataset is used to train a teacher classifier. The teacher classifier labels web data. Finally, both mobile and web data are used to train a student classifier using a noisy training strategy.}
    \caption{We applied semi-supervised learning to boost screen classification performance using unlabeled web data. First, a teacher classifier is trained using a ``gold" dataset of labeled mobile screens. Then, the teacher classifier is used to generate a ``silver" dataset of pseudo-labels by running it on a large, unlabeled data source (\textit{e.g.,} web data). Finally, the ``gold" and ``silver" datasets are combined when training a student classifier, which is larger and regularized with noise to improve generalization. This process can be repeated; however, we only perform one iteration.}
    \label{fig:diagram_ssl}
\end{figure}
Classifying screen type or functionality from a screenshot can be useful for design analysis and automation. Previously, small amounts of data have been collected and annotated for this purpose.
Enrico \cite{leiva2020enrico} is an example of a dataset (1460 samples, subset of Rico \cite{deka2017rico}) where each screenshot is assigned to one of 20 mutually-exclusive design categories.
Because of the dataset's small size, it is challenging to train accurate deep learning classification models.
While our web dataset is large, it also does not have the screen-type annotations, and thus it is not possible to employ the same pre-training strategy that was used for element detection.

Instead, we applied a semi-supervised learning technique known as self-training \cite{chapelle2009semi}.
Self-training is a process that improves model performance by iteratively labeling and re-training on a large source of unlabeled data.
We investigated the effects of using WebUI as the unlabeled dataset and show that doing so improves overall screen classification accuracy.

\subsubsection{Model Implementation}
Figure \ref{fig:diagram_ssl} shows our procedure for incorporating WebUI data into our model training via self-training.

First, we trained screen classifier based on the VGG-16 architecture with batch normalization and dropout \cite{simonyan2014very}, as described by the Enrico paper \cite{leiva2020enrico}.
Since official training, validation, and testing splits were not provided, we randomly generated our own (70\%/15\%/15\%).
This model was trained only on data from the Enrico training split and served as the \textit{teacher classifier}.
Next, the teacher model was used to generate ``soft" pseudo-labels for screenshots in the WebUI dataset, where each sample was mapped to a vector containing probabilities for each class.
We followed the procedure used by Yalniz et al. \cite{yalniz2019billion} to keep only the top K most confident labels for each class.
To select K, we first randomly sampled a small subset of 1000 web pages from our dataset and performed a parameter search to find the optimal value. Based on our experiments, we found that a value of 10\% of the total dataset size led to good performance (\textit{e.g.,} we set K=700 for the Web-7k split).
Finally, we trained a \textit{student classifier} on a combination of the original and automatically generated labels.
We employed a specific type of self-training known as Noisy Student Training \cite{xie2020self}, which involves injecting noise into the student model's training process so that it becomes more robust.
Two types of noise are used in this process: \textit{(i)} input noise, which is implemented via random data augmentation techniques and \textit{(ii)} model noise, which is implemented with dropout \cite{srivastava2014dropout} and stochastic depth \cite{huang2016deep}.
Because stochastic depth can only be applied to model architectures with residual blocks, we used an architecture based on ResNet-50 \cite{he2016deep} instead of VGG-16.
\subsubsection{Results}
\begin{table}[!]
\Description{Classification accuracy for 6 different model configurations. The baseline configuration achieves an accuracy of 0.4737 while our best configuration achieves an accuracy of 0.5263.}
\caption{Classification accuracy (across 20 classes) for different configurations of our screen classification model. Increasing the amount of data used with our semi-supervised learning method led to increased accuracy.}
\label{tab:results_screenclassification}
\begin{tabular}{@{}ll@{}}
\toprule
Model Configuration           & Accuracy \\ \midrule
VGG-16         & 0.4737   \\
Noisy ResNet-50 & 0.4649 \\
Noisy ResNet-50 (Rico)    & 0.4956   \\
Noisy ResNet-50 (Web7k)   & 0.4864   \\
\textcolor{black}{Noisy ResNet-50 (Web7k-Resampled)} & \textcolor{black}{0.4868} \\
Noisy ResNet-50 (Web70k)  & 0.5175   \\
Noisy ResNet-50 (Web350k) & 0.5263   \\ \bottomrule
\end{tabular}
\end{table}

Overall, we found that applying self-training to incorporate additional unlabeled data led to consistent performance improvements (Table \ref{tab:results_screenclassification}).
The best classifier using WebUI data was 5\% more accurate than the baseline model, which was only trained with the Enrico dataset.
Our baseline VGG-16 model performed considerably worse than the originally reported results \cite{leiva2020enrico} but achieved similar accuracy to another reproduction of the work \cite{liang2021multibench}. The performance difference could be attributed to differences in randomized splits. Since we used the same splits across all conditions, we expect relative performance differences to be consistent.
To investigate the effects of using a new model architecture, we also trained a Noisy ResNet-50 (architecture used by the student model) on the Enrico dataset.
The resulting classifier performed relatively poorly (worse than the baseline model), since the modifications introduced (dropout and stochastic depth) require more data to train effectively.

The primary source of improvement stems from the inclusion of additional unlabeled data during the training process, which led to a more generalizable student model.
We observed that the small size of the Enrico dataset (1460 samples) quickly led to overfitting during training and limited overall performance.
Semi-supervised learning techniques, such as self-training, allow training on a much larger number of examples.
We found that model accuracy improved when we incorporated more unlabeled examples, both from WebUI and Rico.

\subsection{\textcolor{black}{Screen Similarity}}
\begin{figure}[!htb]
    \centering
\includegraphics[trim={10pc 10pc 3pc 10pc},clip,width=15pc]{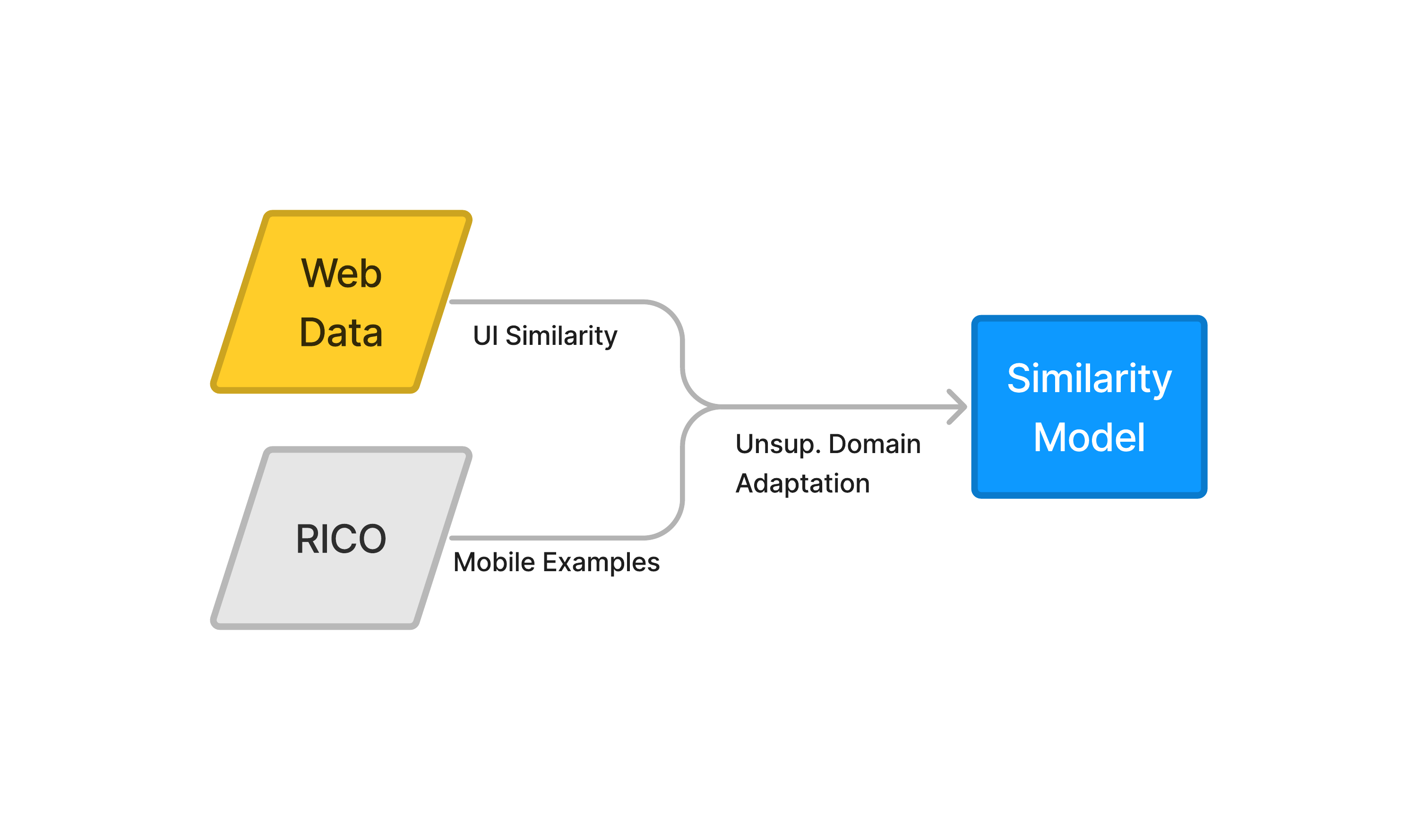}
\Description{Flow chart demonstrating how unsupervised domain adaptation is applied to screen similarity model. Both web data and mobile data are fed simultaneously into a similarity model during training.}
    \caption{We used unsupervised domain adaptation (UDA) to train a screen similarity model that predicts relationships between pairs of web pages and mobile app screens. The training uses web data to learn similarity between screenshots using their associated URLs. Unlabeled data from Rico is used to train an \textit{domain-adversarial} network, which guides the main model to learn features that transferrable from web pages to mobile screens.}
    \label{fig:diagram_uda}
\end{figure}
Identifying variations within the same screen and detecting transitions to new screens are useful for replaying user interaction traces, processing bug reports \cite{cooper2021takes}, and automated app testing \cite{li2017droidbot,li2019humanoid}.
To model these properties and understand how multiple screens from an application relate to each other, previous work \cite{feiz2022understanding,li2019humanoid} has sought to differentiate between distinct UIs and variations of the same UI.
For example, the same checkout screen may appear different based on the number and types of products added to the cart. Common screen interactions such as scrolling and interaction with expandable widgets (\textit{e.g.,} menus, dialogs, keyboards, and notifications) may also alter the visual appearance of a screen.
Visual prediction reduces system reliance on accessibility metadata, which may be missing or incomplete, and further extends the applications of these models, as they can process video recordings of user interactions (\textit{e.g.,} reproducing bug reports) \cite{bernal2022translating, cooper2021takes}.

Previous work \cite{feiz2022understanding} opted to manually annotate a dataset of more than one thousand iPhone applications that were manually ``crawled" by crowdworkers; however, the dataset was not released to the public.
As a weak source of annotation, we used web page URLs to automatically label page relations.
Since no labeled data is available in the mobile domain, we employed domain-adversarial network training \cite{ganin2016domain}, a type of unsupervised domain adaptation (UDA), to encourage the model to learn transferrable features from the web domain that might apply to the mobile domain.
Note that while it is possible to apply the \textcolor{black}{semi-supervised learning} strategy \textcolor{black}{(which was used for the screen classification task)} in reverse, it may be less effective, since the unlabeled dataset (mobile UIs) is smaller than the labeled dataset.
\subsubsection{Model Implementation}
We followed previous work \cite{feiz2022understanding} and used a ResNet-18 \cite{he2016deep} model trained as a siamese network \cite{hadsell2006dimensionality}.
The siamese network uses the same model to encode two inputs, then compares them in feature space (\textit{i.e.,} their embeddings) to decide if they are different variations of the same UI screen.
Our approach is different from the method proposed by previous work \cite{cooper2021takes}, which applies random data augmentations (\textit{e.g.,} blurring, rotation, translation) to screenshots to create \textit{same-screen} pairs.
Instead, we randomly sampled pairs of screenshots from our web data for training, with balanced probability for same-screen and new-screen pairs.
Same-screen pairs were generated by finding screenshots with the same URL but accessed at different times or simulating page scrolls on a full-page screen capture by sliding a window vertically along the image.
Note that occasionally, simulated page scrolls and accessing the same web page at different times still produced identical or nearly identical screenshots, so in our test set, we filtered these out using perceptual hashing.
Different-screen pairs were generated both by sampling screenshots from within the same domain but with different URL path, and by sampling screenshots from other domains.

The domain-adversarial training process seeks to simultaneously accomplish two objectives: \textit{(i)} learn an embedding space where two screenshots are from the same screen if their distance is less than a threshold, and \textit{(ii)} learn an encoding function that applies to both the web and mobile domains.
The first objective is related to the primary task of distinguishing same-screen pairs from new-screen pairs and is achieved with a pairwise margin-based loss \cite{feiz2022understanding}.
The second objective aims to align the feature distributions of the two domains by \textit{maximizing} the error rate of a domain classifier, which is a network that tries to classify whether a sample is from a web or mobile UI.
For this task, we used only web page screenshots captured on simulated smartphones, to make the domain classification objective more challenging.

\subsubsection{Results}
\begin{figure*}[!]
    \centering
\includegraphics[width=\textwidth]{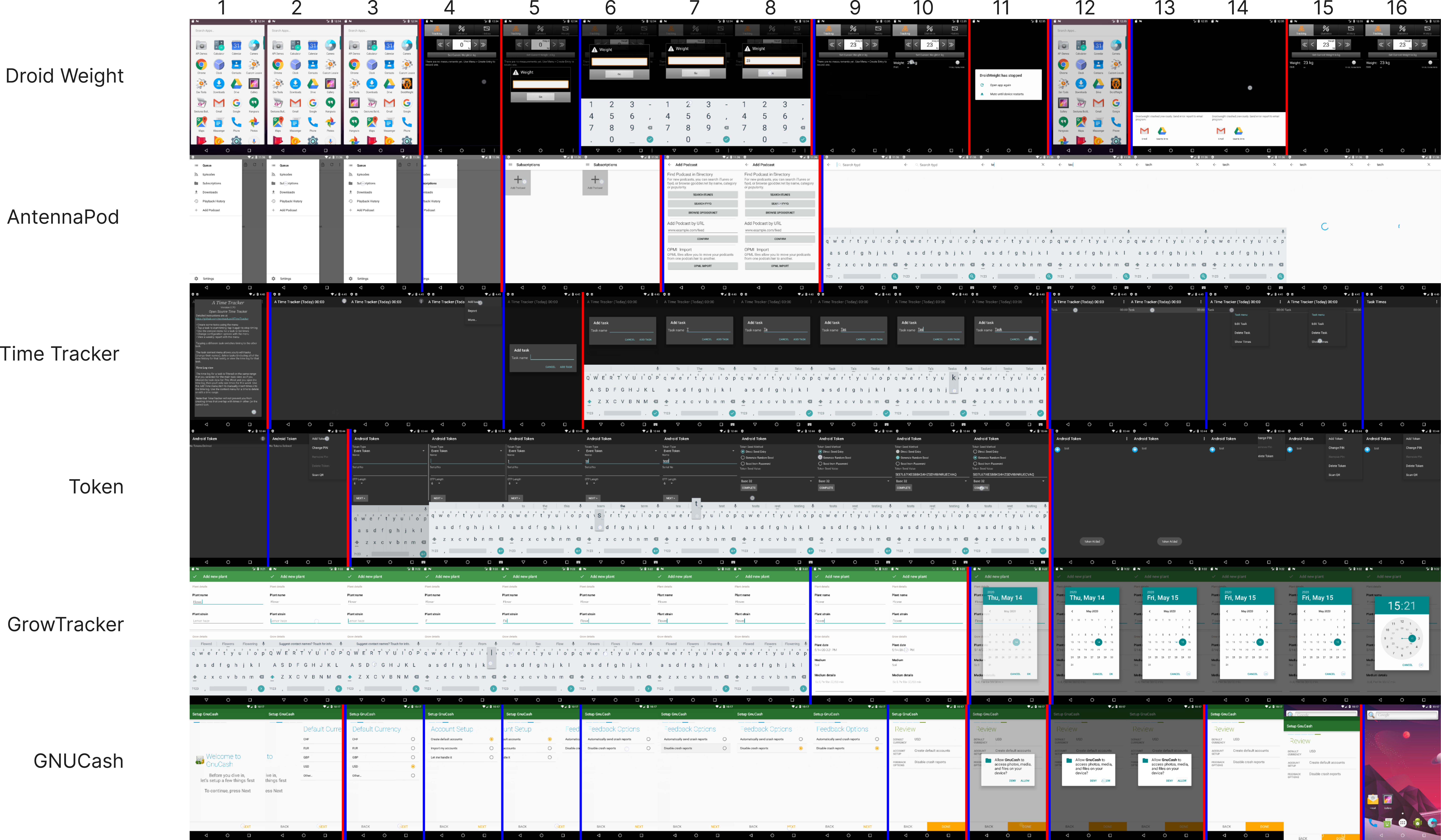}
\Description{6 video timelines processed by our crawl segmentation model. Each video is broken into 16 frames. Lines are drawn where our models predicted that the user entered a new screen. We compared a UDA and non-UDA model and show qualitatively that the UDA model produces more accurate segmentations.}
    \caption{Examples of interaction videos segmented by our best models trained with UDA (Red) and without UDA (Blue). Videos are sampled at 1 fps. The output of both models contain errors, however, we found that the adapted UDA model generally produced better segmentations. Common errors include oversegmentation due to app dialogs and soft keyboards, which do not occur in the WebUI dataset.}
    \label{fig:diagram_crawls}
\end{figure*}
\begin{table}[!]
\Description{Classification performance of our 6 configurations of our crawl segmentation model evaluated on pairs of web pages. The lowest performing model was trained on 7k web pages and achieved an F-1 score of 0.7097. The best performing model was trained on 350k web pages and achieved an F1-score of 0.9630.}
\caption{Classification performance (\textit{same-screen} vs \textit{new-screen}) of our \textcolor{black}{screen similarity} models evaluated on pairs of screens from our web data. Performance increased when the model was trained on more data and slightly decreased when trained with the UDA objective.}
\label{tab:table_screensegment}
\begin{tabular}{ll}
\hline
Model Configuration     & F1-Score \\ \hline
ResNet-18 (Web7k)       & 0.7097   \\
ResNet-18 UDA (Web7k)   & 0.7184   \\
\textcolor{black}{ResNet-18 (Web7k-Resampled)} & \textcolor{black}{0.7368} \\
\textcolor{black}{ResNet-18 UDA (Web7k-Resampled)} & \textcolor{black}{0.7191} \\
ResNet-18 (Web70k)      & 0.8222   \\
ResNet-18 UDA (Web70k)  & 0.8193   \\
ResNet-18 (Web350k)     & 0.9630   \\
ResNet-18 UDA (Web350k) & 0.9500   \\ \hline
\end{tabular}
\end{table}
Since one of the assumptions of our problem is that labeled examples of same-screen and new-screen pairs are unavailable for mobile apps, we used two alternative methods to evaluate our \textcolor{black}{screen similarity} model: \textit{(i)} quantitative evaluation on labeled pairs of web screens and \textit{(ii)} qualitative evaluation on a set of unlabeled Android interaction videos.

Table \ref{tab:table_screensegment} shows the quantitative performance of our models evaluated on pairs of web pages from our dataset.
Overall, training with more data led to significantly better performance, an increase of over 20\%.
The inclusion of a domain adaptation objective sometimes led to a slight drop in classification performance since it introduces additional constraints in the learning process.
We qualitatively evaluated our model's performance characteristics on mobile screens by using them to segment videos of mobile app interaction.
We used a dataset of screen recordings of bug reproductions \cite{cooper2021takes} for 6 open-source Android apps and applied our model by sequentially sampling frames from the video and evaluating whether a new screen was reached.
Note our sampling process differs from other previous work \cite{deka2017rico,burns2022interactive} that segmented crawls at recording time using accessibility metadata, because we do not have accessibility metadata corresponding to the previously collected recordings used in our analysis.
Figure \ref{fig:diagram_crawls} shows an example of a usage video processed by our model.
While the web model was effective detecting some types of transitions that occurred in mobile apps, it was less effective at others, such as software keyboards and dialogs, which do not occur frequently in the WebUI dataset.
We include more model-generated segmentations of the bug reproduction dataset in supplementary material.

In this work, we applied \textit{unsupervised} domain adaptation, which does not require any labels from the target domain. Other domain adaptation strategies exist, and some are able to incorporate small amounts of labeled data, which we expect could improve the accuracy of our model by contributing transition types unique to mobile apps.
\section{Discussion}

\subsection{Performance Impact of Web Data}

Empirically, we showed that automatically crawled and annotated web pages, like those available in WebUI, can effectively support common visual modeling tasks for other domains (\textit{e.g.,} mobile apps) through transfer learning strategies.
In cases where a small amount of labeled mobile data was available, as in element detection and screen classification, incorporating web data led to better performance.
Even when labeled data was completely unavailable, as in \textcolor{black}{screen similarity}, models trained only on web data could often be directly applied to mobile app screens.
Our results suggest that the size of current UI datasets may be a limiting factor, since model performance increases consistently when trained on larger splits of data. 
Our observations and analysis of WebUI's composition showed that web pages can differ from mobile app screens in terms of complexity (\textit{i.e.,} average number of on-screen elements) and element types.
However, the performance improvements from our machine learning experiments suggest that web and mobile UIs are similar enough to transfer some types of semantics between them.

We currently only explored three examples, although we believe that other UI modeling works \cite{wu2021screen,chen2022towards,swearngin2019modeling} can also benefit from similar approaches. %
We did not evaluate all possible applications of WebUI in our paper, due to time and cost constraints.
However, the three experiments we conducted cover all possibilities of source and target domain labels (\ref{tab:strategies}), so similar transfer learning techniques are likely to apply.
Future work that builds upon WebUI can conduct more detailed evaluations of other downstream tasks.

One specific area that we believe is promising for future work is automated design verification \cite{moran2018automated}, which could benefit from a large volume of web pages containing paired visual and stylistic information.
Our highly automated data collection process also allows WebUI to be more easily updated in the future by re-visiting the same list of URLs.
An updated version of the dataset could also facilitate longitudinal analysis of the design \cite{deka2021early} and accessibility \cite{fok2022large} of web UIs.
Nevertheless, WebUI is currently unlikely to support other types of modeling, such as user interaction mining \cite{deka2016erica,deka2017rico}, that require realistic interaction traces, since our crawling strategy was largely based on random link traversal.

\subsection{Improved Automated Crawling}
Our crawler was unable to access much of the ``deep web" (\textit{i.e.,} large part of the web that cannot be indexed), and thus our dataset contains few, if any, web pages that are not publicly accessible or protected by authentication flows.
It also did not attempt to interact with all elements on a web page and conducted a very limited exploration of any JavaScript-enabled functionality that might have been present.
Trends in web and app development, such as the creation of Progressive Web Apps (PWAs), suggest that this type of functionality will become more common, and traditional link-based traversal may become less effective at exploring UI states.

To improve automated crawling and data collection, our crawler could benefit from a semantic understanding of web pages.
For example, it could detect page functionality to explore states that require human input and either execute automated routines (\textit{e.g.} detecting login fields) or employ crowdsourcing \cite{deka2017rico} to allow it to proceed in more complex scenarios.
Our currently trained models could augment or improve this process by identifying tasks associated with web pages (\textit{e.g.,} screen classification) or by augmenting potentially noisy labels provided by the automatically generated accessibility tree.
In turn, the crawler could explore more of the web, leading to higher quality and more diverse data.
If repeated iteratively, this process would constitute a form of Never-Ending Learning \cite{mitchell2018never}, a machine learning paradigm where models learn continuously over long periods of time.
Instead of learning from a fixed dataset, models could constantly improve itself by encountering new content and designs, both of which are important due to the dynamic nature of UIs.
\subsection{Generalized UI Understanding}
Our experiments show that incorporating web data is most effective for improving visual UI modeling in transfer learning settings where a limited amount of target labels are available for fine-tuning.
A logical next step is to obtain similar benefits without any additional labeled data.
To this end, we identified several strategies for improving generalization.
First, unlike existing UI datasets that contain examples from one device type, we intentionally simulated multiple viewports and devices during data collection.
The decomposition of one-hot labels \textcolor{black}{(where each element type is assigned exactly one type)} into combinations of multi-hot tags \textcolor{black}{(each element can be assigned multiple labels)} may also be useful, since it avoids the problem of platform-specific element types.
Figure \ref{fig:diagram_zsl} demonstrates the zero-shot transfer capabilities of models trained only on web data by successfully detecting and classifying elements on Android app screens.
While the label sets of web and Android data do not directly overlap, the web model outputs reasonable analogs (\textit{e.g.,} Text, link) for Android widgets (\textit{e.g.,} Text Button).
Finally, our \textcolor{black}{screen similarity} model shows how \textit{unsupervised domain adaptation} can improve the transferrability of learned features across domains through an explicit machine learning objective.

A long-term goal of our automated data collection and modeling efforts is achieving a more generalized understanding of UIs --- a single model that could be used to predict semantics for any UI.
This is challenging due to differing design guidelines and paradigms, but it could ultimately lead to a better understanding of how to solve UI problems across platforms.

\section{Conclusion}
In this paper, we introduced WebUI, a dataset of approximately 400,000 web pages paired with visual, semantic, and style information to support visual UI modeling.
Unlike most existing datasets for UI research that depend on costly and time-consuming human exploration and annotation, WebUI was collected with a web crawler that uses existing metadata, such as the accessibility tree and computed styles, as noisy labels for visual prediction.
Our highly automated process allowed us to collect an order of magnitude more UIs than other publicly released datasets and often associates more information (\textit{e.g.,} clickability, responsiveness) with each example.
We demonstrated the utility of our dataset by incorporating it into three visual UI modeling tasks in the mobile domain: \textit{(i)} element detection, \textit{(ii)} screen classification, and \textit{(iii)} \textcolor{black}{screen similarity}.
In cases where a small amount of labeled mobile data exists, incorporating web data led to increased performance, and in cases without any labeled mobile data, we found that models trained on web pages could often generalize to mobile app screens.
In summary, our work shows that the web constitutes a large source of data that can more sustainably be crawled and mined for supporting visual UI research and modeling. 

\begin{acks}
This work was funded in part by an NSF Graduate Research Fellowship.
\end{acks}

\bibliographystyle{ACM-Reference-Format}
\bibliography{main}

\appendix

\begin{figure*}[!]
    \centering
\includegraphics[width=0.9\textwidth]{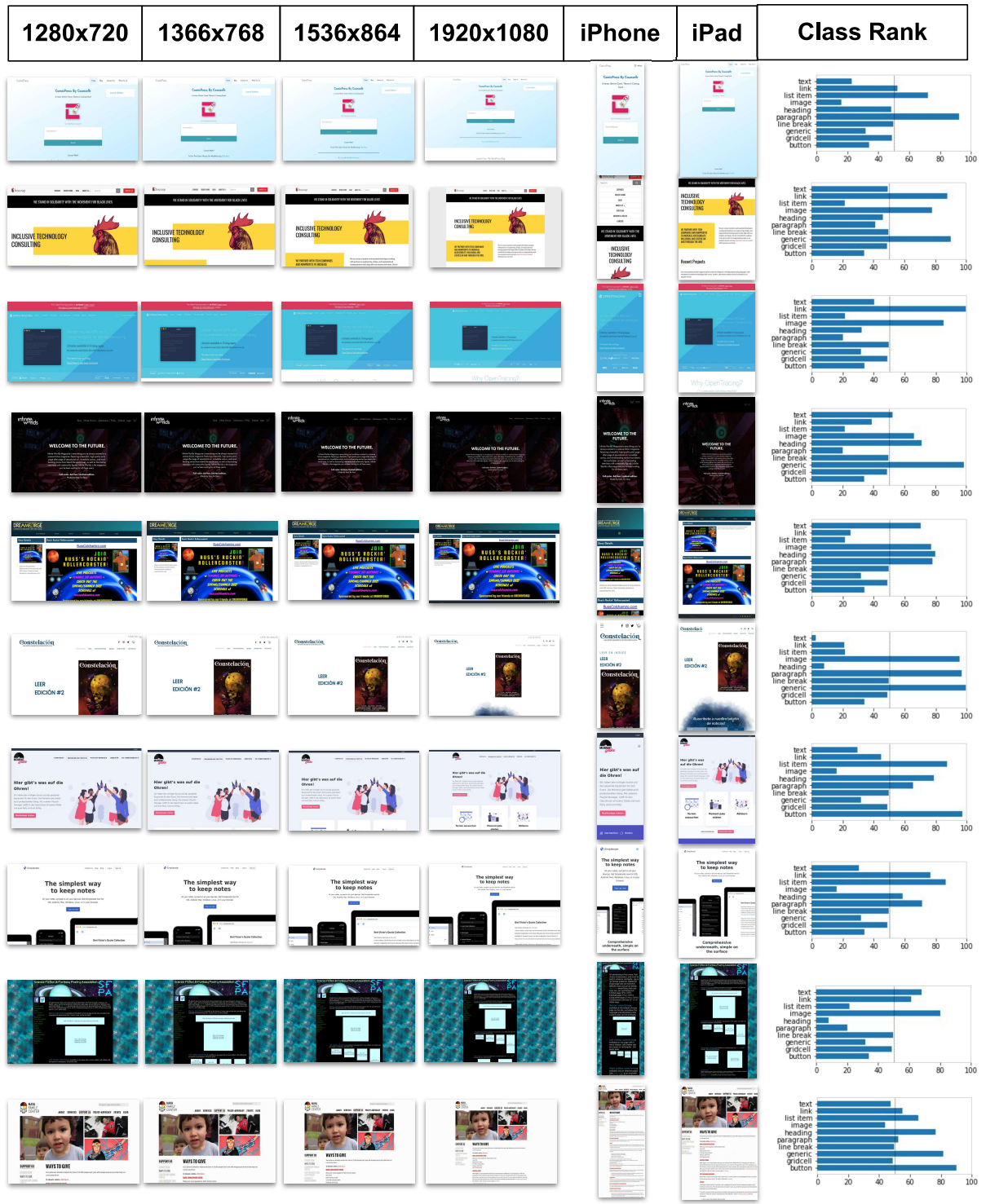}
\Description{A gallery of 10 websites rendered using 6 simulated devices each. The gallery contains a variety of web pages such as login pages, landing pages, product pages, and portfolios. Each page is characterized by a class rank bar chart which characterizes the relative frequency of its element classes.}
    \caption{\textcolor{black}{Samples from WebUI accessed with different simulated devices. For each screen, we compute its element type distribution (normalized to 1). Then, we computed the percentile-rank of the top 10 classes with respect to the entire dataset. For example, the bottom row's \texttt{button} class has a percentile-rank of 90, meaning the web page's relative frequency of is greater than 90\% of others in the dataset.}}
    \label{fig:collage1}
\end{figure*}

\textcolor{black}{\section{Additional Dataset Samples}
We provide additional samples from the WebUI (Figure \ref{fig:collage1}) to supplement the example in the paper (Figure \ref{fig:diagram_devices}). Our example gallery shows several different types of websites, including login, landing, product, portfolio, and informational pages. Each website is captured using different simulated devices, which shows, among other things, how content responds to screen size. We also computed the percentile-rank of each web page's class distribution.}

\textcolor{black}{
\section{Class Imbalance Analysis}
This section describes analysis of class imbalance of WebUI and its effect on transfer learning applications.
Similar to other UI datasets\cite{zhang2021screen}, WebUI exhibits an imbalance of UI element classes, where some types of elements (\textit{e.g.,} text) appear much more frequently than others (\textit{e.g.,} images).
Several aspects of WebUI (\textit{e.g.,} finer-grain text segmentation, multi-hot labels, and prevalence of documents on the web) also contributed to class imbalance.
}

\textcolor{black}{We used a frequency-based resampling method to generate the Web7k-Resampled, which resulted in more examples of infrequent element types.
Our technique assigned weights to samples to increase the representation of UIs containing rare or infrequent element types, and we resampled based on the 10 element types shown in Figure \ref{fig:diagram_composition}.
Algorithm \ref{alg:resampling} provides an overview of our resampling technique.
Note that unlike some class-balancing algorithms (\textit{e.g.,} SMOTE \cite{chawla2002smote}), our technique does not generate additional synthetic samples and does not include the same screen more than once.
}

\begin{figure}[!]
    \centering
\includegraphics[width=\linewidth]{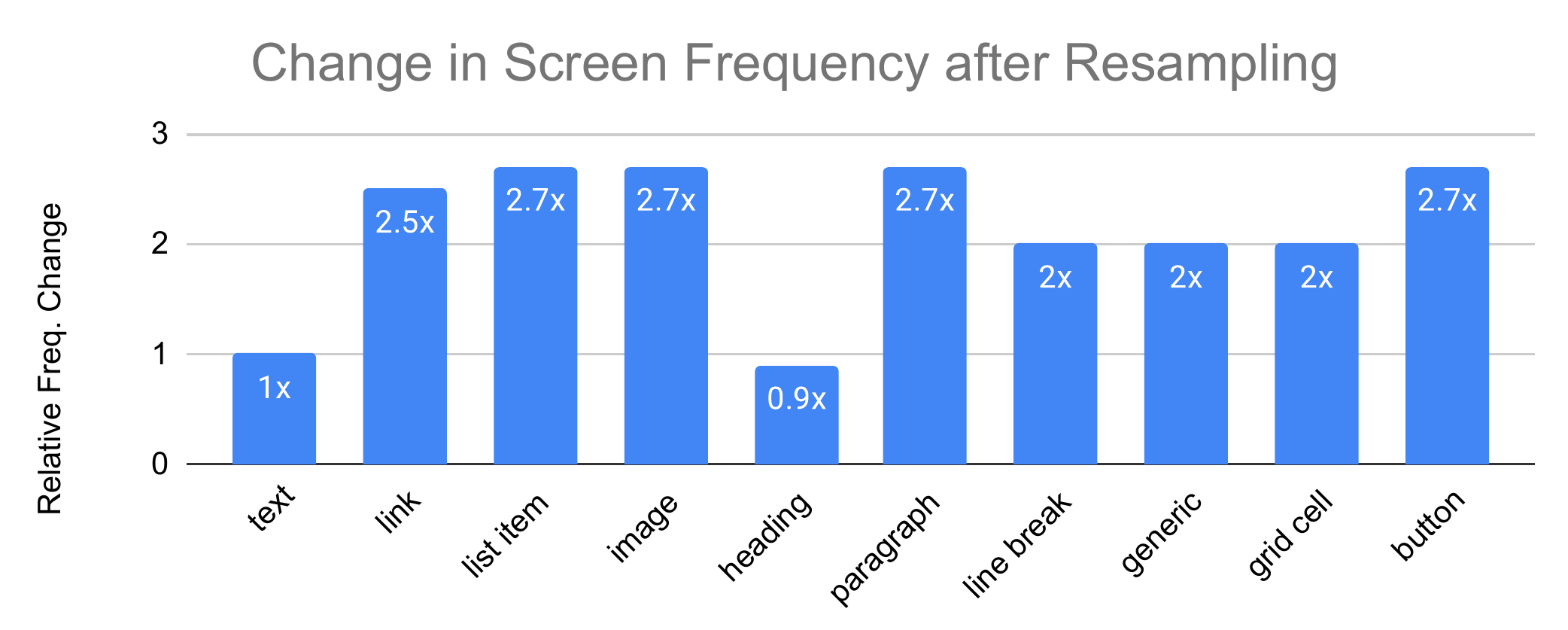}
\Description{A bar chart showing the change in frequency of screens containing at least one of each element type after resampling. With the exception of text and headings, all classes have more than 2 times more screens in the resampled split.}
    \caption{\textcolor{black}{We calculated the change in frequency (expressed as a ratio) of screens containing at least one of each element type after resampling. For example, the number of screens containing at least one image element is 2.7x more than in the randomly sampled set.}}
    \label{fig:resample_screen}
\end{figure}

\begin{figure}[!]
    \centering
\includegraphics[width=\linewidth]{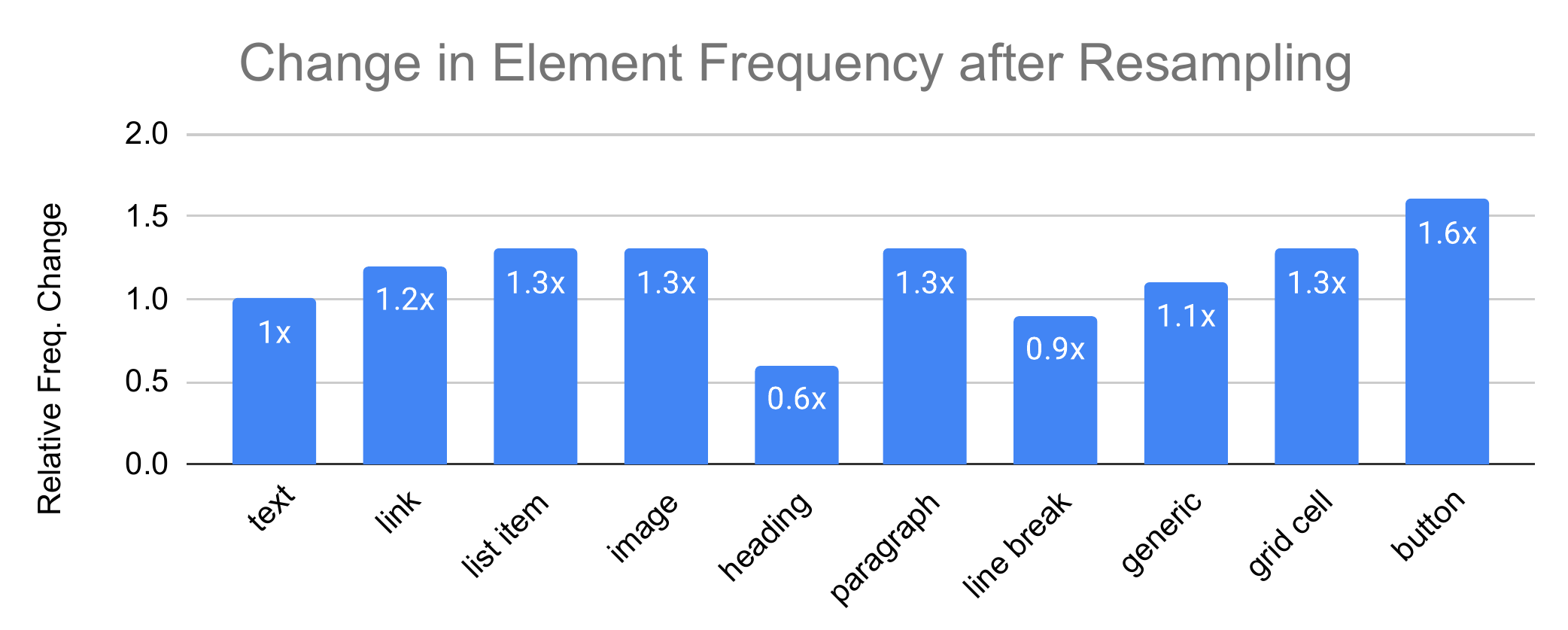}
\Description{A bar chart showing the change in frequency of elements after resampling. Most elements have increased representation, of around 1.1 to 1.3 times more.}
    \caption{\textcolor{black}{We calculated the change in frequency (expressed as a ratio) of total number of elements after resampling. For example, the average screen in the resampled split contains 1.3x more images. Note that is possible for most element classes to increase in frequency (while not having other classes experience a proportional decrease) because element classes are not mutually exclusive, and the resampled split contains more elements that are assigned multiple tags.}}
    \label{fig:resample_element}
\end{figure}

\textcolor{black}{Web7k-Resampled contains proportionally more examples of many infrequent classes (Figure \ref{fig:diagram_composition}).
Figure \ref{fig:resample_screen} shows the proportional increase in screens containing each element type.
Figure \ref{fig:resample_element} shows the proportional increase in the total number of elements for each type.
}

\begin{table*}[!]
\small
\centering
\Description{Table containing the average precision of each element class in the Element Detection class. Numbers represent a breakdown of the previously reported mAP value and we show that class balancing has only a small effect on the finetuned model accuracy.}
\caption{\textcolor{black}{Average Precision (AP) of each element class (excluding the ``Other" class) for the Element Detection task.}}
\label{tab:ap_breakdown}
\begin{tabular}{@{}lllllll@{}}
\toprule
Element Type     & SSD (Random) & FCOS (Random) & FCOS (Web7k) & FCOS (Web7k-Re.) & FCOS (Web70k) & FCOS (Web350k) \\ \midrule
Background Image & 0.85         & 0.88          & 0.86    &0.91     & 0.85          & 0.93           \\
Checked View     & 0.06         & 0.28          & 0.31    &0.34     & 0.32          & 0.38           \\
Icon             & 0.72         & 0.73          & 0.75    &0.75     & 0.75          & 0.77           \\
Input Field      & 0.22         & 0.59          & 0.7     &0.60     & 0.72          & 0.69           \\
Image            & 0.73         & 0.8           & 0.77    &0.82     & 0.78          & 0.82           \\
Text             & 0.66         & 0.83          & 0.89    &0.84     & 0.9           & 0.85           \\
Text Button      & 0.57         & 0.9           & 0.94    &0.94     & 0.95          & 0.94           \\
Page Indicator   & 0.83         & 0.76          & 0.83    &0.76     & 0.79          & 0.8            \\
Pop-Up Window    & 0.85         & 0.83          & 0.8     &0.85     & 0.78          & 0.83           \\
Sliding Menu     & 0.95         & 0.98          & 0.96    &0.98     & 0.96          & 0.97           \\
Switch           & 0.97         & 0.93          & 0.86    &0.97     & 0.91          & 0.94           \\ \midrule
mAP              & 0.67         & 0.77          & 0.79    &0.80     & 0.79          & 0.81          
\end{tabular}
\end{table*}
\begin{algorithm}
    \SetKwInOut{Input}{Input}
    \SetKwInOut{Output}{Output}

    \underline{function SampleSplit} $(N, C, S)$\;
    \Input{Number of samples to choose $N$, list of element classes $C$, and list of samples $S$}
    \Output{Resampled subset of $S$}

    \tcc{Vector containing total frequencies for $c \in C$}
    $f_C \leftarrow$ total \# of elements in $S$ for each class

    \tcc{Matrix where rows are $s \in S$ and columns are normalized frequency of $c \in C$ for $s$}
    $f_S \leftarrow$ frequency of classes $c \in C$ (columns) for $s \in S$ (rows)

    \tcc{Assign sampling weights to $c \in C$ inversely proportional to frequency}
    $w_C \leftarrow$ [$\frac{1}{f_C[c]}$ | $c \in C$]
    
      $\mathrm{samples} \leftarrow \mathrm{[]}$

      \tcc{Repeat until desired split size is reached}
      \While{$\mathrm{\mathbf{len}(samples)} < N$}{
      $c_s \leftarrow$ $\mathbf{Sample}(C, w_C)$

      $w_s \leftarrow$ [$f_S[s, c_s]$ | $s \in S$]

      sample $\leftarrow$ $\mathbf{SampleWithoutReplace}(S, w_s)$

        add sample to samples
      }
      return samples
        \caption{\textcolor{black}{Pseudo-code for the frequency-based resampling algorithm used to generate the Web7k-Resampled split.}}
    \label{alg:resampling}
\end{algorithm}
\textcolor{black}{The results from our performance evaluations in the main paper suggest that this resampled split leads to improvements for each of our three tasks when compared to a randomly sampled subset of the same size.
Notably, the element detector model resampled 7k split outperformed the one trained on 70k random split, which suggests that element balancing was particularly useful for tasks where elements types are directly predicted.
Tests with other two tasks (screen classification and screen similarity) also led to improvements for the resampled models; however, the gains were more modest.
The improvements could be because the element distribution in the resampled split is closer to that of the target data.
In addition, we provide a deeper analysis of the Element Detection class, which is most likely to be affected by element type imbalance. Table \ref{tab:ap_breakdown} shows that the Web7k-resampled split has higher AP for classes like "Text Button" and "Image", which had increased representation after resampling.}

\end{document}